\documentclass{jfm}
\usepackage{soul}
\usepackage{graphicx}
\usepackage{newtxtext}
\usepackage{newtxmath}
\usepackage{natbib}
\usepackage{hyperref}
\usepackage{soul, color, xcolor}
\hypersetup{
	colorlinks = true,
	urlcolor = blue,
	citecolor = black,
}

\newcommand{\RomanNumeralCaps}[1]
\linenumbers

\soulregister{\cite}7 

\usepackage{soul}
\soulregister\cite7
\soulregister\ref7 
\soulregister\citep7 
\soulregister\citet7 
\soulregister\pageref7 

\title{Anomalous Reynolds stress and dynamic mechanisms in two-dimensional elasto-inertial turbulence of viscoelastic channel flow}
\author{Haotian Cheng\aff{1}, Hongna Zhang\aff{1}\corresp{\email{hongna@tju.edu.cn}} , Wenhua Zhang\aff{1}\corresp{\email{zhangwh2022@tju.edu.cn}}, Suming Wang\aff{1}, Xiaobin Li\aff{1}, Yuke Li\aff{2}\and Fengchen Li\aff{1}}

\affiliation{\aff{1}State Key Laboratory of Engine, Tianjin University, Tianjin 300350, PR China \aff{2}College of Shipbuilding Engineering, Harbin Engineering University, Harbin 150001, PR China and National Key Laboratory of Ship Structural Safety, Harbin 150001, Heilongjiang, PR China}
\begin{document}
	\maketitle
	
	\begin{abstract}
	Elasto-inertial turbulence (EIT) has been demonstrated to be  able to sustain in two-dimensional (2D) channel flow; however the systematic investigations on 2D EIT remain scare. This study addresses this gap by examining the statistical characteristics and dynamic mechanisms of 2D EIT, while exploring its similarities to and differences from three-dimensional (3D) EIT. We demonstrate that the influence of elasticity on the statistical properties of 2D EIT follows distinct trends compared to those observed in 3D EIT and drag-reducing turbulence (DRT). These differences can be attributed to variations in the underlying dynamical processes. As nonlinear elasticity increases, the dominant dynamic evolution in 3D flows involves the gradual suppression of inertial turbulence (IT). In contrast, 2D flows exhibit a progressive enhancement of EIT. More strikingly, we identify an anomalous Reynolds stress in 2D EIT that contributes negatively to flow resistance, a behavior opposite to that of IT. Quadrant analysis of velocity fluctuations reveals the predominance of motions in the first and third quadrants. These motions are closely associated with polymer sheet-like extension structures, which are inclined from the near-wall region toward the channel center along the streamwise direction. Finally, we present the dynamical budget of 2D EIT, which shows significant similarities to that of 3D EIT, thereby providing compelling evidence for the objective existence of the 2D nature of EIT.
	\end{abstract}
	
	\section{Introduction}
	\label{sec:intro}	
    Viscoelastic fluids are ubiquitous in industrial production and biological fluids and exhibit distinctive flow behaviors, especially  turbulence drag reduction (TDR) at high Reynolds number (Re, \cite{Toms49}) and the elastic turbulence (ET) at extremely low Re (Grossman (2001)). The early turbulence (\cite{Ostwald26}) and the maximum drag reduction (MDR) asymptote of TDR (\cite{Virk1970}) in viscoelastic fluid have long been mysteries. The discovery of elasto-inertial turbulence (EIT) by \cite{Samanta13} and \cite{Dubief13} provides a harmonizing explanation for both the early turbulence and the MDR phenomena, opening up a new avenue for viscoelastic turbulence.
        
    Unlike elastic turbulence (ET) at extremely low Re and inertial turbulence (IT), elasto-inertial turbulence (EIT) is primarily governed by elastic nonlinearity with non-negligible inertial nonlinearity. Consequently, the structure and self-sustaining mechanisms of EIT are closely linked to elastic instability (\cite{Dubief23}). The friction factor of EIT follows the MDR asymptote and exhibits characteristics consistent with the MDR state, such as a mean velocity profile approaching the Virk asymptote and negligible Reynolds stress (\cite{Warholic99}; \cite{Min03}). Recent studies have revealed that the flow topology of EIT differs significantly from that of IT. While IT is typically characterized by small-scale hairpin vortex structures, EIT is marked by localized, strongly elongated sheet-like structures formed by polymer extension in the streamwise direction, accompanied by alternating spanwise cylindrical structures (\cite{Dubief13, Dubief23}; \cite{Terrapon15}). The formation of these unique structures is attributed to the energy transfer mechanism from polymer stress work fluctuations to turbulent kinetic energy (\cite{Dubief13}) and the role of polymeric pressure fluctuations (\cite{Terrapon15}). Additionally, experimental studies have identified the characteristic chevron-shaped streak structures of EIT, which transition into inclined near-wall streaks and shear layers as inertia increases (\cite{Choueiri21}). Regarding the self-sustaining mechanism of EIT, we conducted systematic analyses by incorporating EIT-related dynamics into drag-reducing turbulence (DRT) (\cite{Zhang22}). This approach allowed us to redefine the self-sustaining cycle of DRT, where IT-related and EIT-related dynamics coexist. Our findings reveal that IT-related dynamics is gradually supplanted by EIT-related dynamics in DRT as elasticity increases (\cite{Zhang22}). Despite these advancements, the unique dynamic mechanisms underlying EIT remain an active area of research (\cite{Dubief23}).	
	
	The delayed recognition of elasto-inertial turbulence (EIT) stems partly from its overlap with inertial turbulence (IT) across a broad parameter space (\cite{Dubief23}). Historically, the maximum drag reduction (MDR) state was interpreted through the lens of polymer-modulated IT, leading some studies (\cite{Xi10a,Xi10b,Xi16}; \cite{Graham14}; \cite{Wang17}) to classify it as a marginal IT state.  Recent advancements have shifted focus toward EIT-dominated flows as the prevailing paradigm. A compelling experimental observation—complete relaminarization of IT within specific polymer concentration ranges at fixed Re (\cite{Choueiri18})—validates the distinct existence of EIT and confirms that the MDR state in viscoelastic drag-reducing turbulence (DRT) fundamentally represents EIT. This relaminarization phenomenon has been corroborated numerically (\cite{Zhang22}). However, our recent findings reveal that the MDR state’s nature is intrinsically tied to the polymer’s maximum extension length $L^2$: while the MDR state of FENE-P fluids with small $L^2$  corresponds to marginal IT, it transitions to EIT-dominated behavior at larger $L^2$ values (\cite{Wang23}). Notably, \cite{Sid18} identified 2D elasto-inertial instability and obtained the sustained EIT in 2D direct numerical simulations (DNS), mirroring the existence of elastic turbulence (ET) in 2D periodic Kolmogorov flow (\cite{Berti08}). Recent experimental work (\cite{Warwaruk24}) further demonstrates that viscoelastic flows exhibit more pronounced 2D straining motions compared to conventional IT, strongly supporting the hypothesis of an inherently 2D nature in EIT. While EIT in 3D DNS and real-world flows necessarily manifests in three dimensions, its interplay with IT complicates targeted investigations. Crucially, 2D DNS offers a unique advantage by effectively isolating IT, thereby enabling dedicated studies of EIT while substantially reducing computational demands.
	
   Subsequently, the study of elastoinertial turbulence (EIT) has experienced rapid expansion through 2D DNS, owing to their relatively low grid requirements and computational efficiency. This computational accessibility has enabled comprehensive investigations into EIT's fundamental characteristics, intrinsic nature, transitional mechanisms, and underlying origins. Notable contributions include the work of \cite{Gillissen19}, who performed 2D DNS of viscoelastic fluid flows across varying polymer concentrations, establishing a classification system distinguishing strongly and weakly coupled EIT regimes based on structural features. A critical divergence between 2D and 3D flow dynamics was revealed by \cite{Zhu21}, demonstrating that unlike three-dimensional flows where friction factors converge to the MDR asymptote with increasing Weissenberg number (Wi), their 2D counterparts exhibit no such convergent behavior.The field advanced significantly with the identification of dual instability mechanisms governing EIT inception: wall-mode instability mediated by Tollmien-Schlichting waves (\cite{Shekar19,Shekar20,Shekar21}) and center-mode instability driven by nonlinear elastoinertial traveling waves (\cite{Page20}; \cite{Choueiri21}). These distinct modes are spatially differentiated, with wall-mode disturbances concentrating near boundary surfaces and center-mode perturbations localizing around flow domain midlines (\cite{Chaudhary19}). Contemporary research consensus, however, increasingly supports wall-mode dominance in EIT generation, as evidenced by recent studies (\cite{Zhang24}; \cite{Beneitez24a}; \cite{Shekar21}). This perspective gains further validation from the discovery of polymer diffusive instability (PDI), which has been classified as a wall-mode phenomenon (\cite{Couchman24}; \cite{Beneitez24b}). Our numerical visualization studies provide direct evidence of wall-mode-induced EIT dynamics, revealing that the cyclical fracture and regeneration of polymer sheet-like structures constitutes the fundamental self-sustenance mechanism (\cite{Zhang24}). Crucially, we have identified the minimal flow unit (MFU) necessary for EIT initiation and maintenance by 2D DNS, establishing a critical foundation for developing numerical criteria for EIT identification. This work further quantifies the Wi-dependent scaling of MFU dimensions while elucidating the underlying physical rationale.
	
	The structural parallels between two- and three-dimensional flows, coupled with the absence of conventional Newtonian near-wall structures, led \cite{Sid18} to propose that EIT becomes the dominant regime in MDR state at elevated Wi. This interpretation implies that the 3D MDR state represents a hybrid regime rather than pure EIT. Fundamental differences emerge between 2D and 3D flows due to the coexistence of inertial turbulence (IT) in three-dimensional configurations, manifesting in divergent flow dynamics and parameter response characteristics. Recent comparative studies by \cite{Zhu21} revealed contrasting Wi-dependent trends in friction factor evolution: while 3D flows exhibit asymptotic convergence toward MDR levels, their 2D counterparts demonstrate non-convergent behavior. Notably, this 3D convergence does not imply stabilization into pure 2D EIT dynamics, as underlying flow structures continue evolving even at high Wi. This conceptual framework positions the 3D MDR state not as a pure EIT analog, but rather as a composite regime incorporating multiple dynamical patterns sustained through elastic effects. \cite{Zhu21} further identified three distinct dynamical cycles within the MDR state, with 2D EIT representing just one phase in this spectrum. Despite significant advances in 2D EIT characterization (\cite{Sid18}; \cite{Shekar21}; \cite{Zhu21}; \cite{Dubief22}), critical gaps persist in systematic quantification of its statistical signatures and underlying dynamical mechanisms.

      To address these knowledge gaps, the present study implements 2D DNS of viscoelastic channel flow using the finitely extensible nonlinear elastic-Peterlin (FENE-P) model. Our methodology adapts analytical frameworks originally developed for 3D DRT and EIT, enabling direct comparison with established 3D flow characteristics. Through this approach, we aim to: (1) establish comprehensive statistical descriptors for 2D EIT, (2) elucidate its unique dynamical processes, and (3) systematically contrast its features with 3D counterparts. The paper is structured as follows: Section \ref{2} details the governing equations and numerical methodology, Section \ref{3} presents comparative analyses of flow statistics and dynamical mechanisms, with concluding discussions provided in Section \ref{4}.
	
	\section {Numerical methodology}\label{2}
	\subsection {Governing equations}
	This study focuses on the 2D plane Poiseuille flow of imcompressible viscoelastic fluid. The channel walls are considered to be non-slip, and the periodic boundary condition is imposed in the streamwise direction. The normalization is performed using the channel half-height $h$, the bulk average velocity $u_b$, the time $h/u_b$, and the pressure $\rho u_b^2$ as characteristic scales of length, velocity, time and pressure. Here, $u_b = \frac{1}{2h}\int_{-h}^{h} \overline{u}(y)$dy with $ \overline{u}(y)$ the locally averaged velocity in the streamwise direction; $\rho$ is the density of fluid. The dimensionless governing equations are as follows:

	\begin{equation}
		\nabla \cdot \textbf{u} = 0\label{EQ.1},
	\end{equation}
	
	\begin{equation}
		\frac{\partial \textbf{u}}{\partial t} + \textbf{u} \cdot \nabla \textbf{u} = - {\nabla p}+ \frac{\beta}{{{\mathop{Re}\nolimits} }} {\nabla^2 \textbf{u}} + \nabla \cdot \boldsymbol{\tau_p} \label{EQ.2},
	\end{equation}
    where $\textbf{u} = (u, v)$ is the velocity vector; $p$ is the pressure; $\boldsymbol{\tau_p}$ is the additional elastic stress tensor;  $\beta = \eta/\mu$ is the the viscosity ratio  with $\mu$ the dynamic viscosity of solution and $\eta$ the contribution of solvent to the viscosity; $Re = \frac{\rho u_bh}{\eta}$ is the average Reynolds number.
		
	The viscoelastic fluid is described by Peterlin (FENE-P) model, based on which $\boldsymbol{\tau_p}$ is obtained by:
	\begin{equation}
		\boldsymbol{\tau_p} =\frac{1-\beta}{Re Wi}[ f(r^2)\textbf{C} - \textbf {I}]\label{EQ.3},
	\end{equation}
	
	\begin{equation}
		\frac{\partial \textbf{C}}{{\partial t }}+(\textbf{u} \cdot \nabla)\textbf{C}- \textbf{C} \cdot (\nabla \textbf{u})- (\nabla \textbf{u})^{\rm T} \cdot \textbf{C}=-\frac{f(r^2)\textbf{C} - \textbf {I}}{\rm Wi}\label{EQ.4},
	\end{equation}
where  $\textbf{C}$ is the conformation tensor indicating the local state of the polymer solution with $\textbf{C}=\langle \textbf{q} \textbf{q}\rangle $ and $\textbf{q}$ the end-to-end vector of a pair of polymer molecules; $Wi = \frac{\lambda u_b}{h}$ is the Weissenberg number;  $\lambda$ is the relaxation time of viscoelastic fluid; the Peterlin function $f(r^2)$ is defined as $f(r^2)$ = $\frac{L^2 - 2}{L^2 - r^2}$ with $r^2$ the trace of $\textbf{C}$ and $L$ the maximum extension length of the polymer chains. 
	
	\subsection {Numerical schemes}
	The above equations are solved by an in-house DNS code based on the staggered-grid finite difference method. For the spatial discretization, the second-order central difference scheme is adopted for the convection, pressure, diffusion, and elastic stress terms of the momentum conservation equation. For the time marching, a fractional step method is employed, where the pressure term uses an implicit scheme and other terms adopts the second-order Adams–Bashforth scheme. A time-splitting method is adopted to solve the governing equations, briefly described as follows: (1) calculating the conformation tensor and the elastic stress field by Eqs. (\ref{EQ.3}) and (\ref{EQ.4}); (2) conducting a partial time marching of the velocity field in Eq. (\ref{EQ.2}) to obtain the first intermediate velocity field; (3) substituting the first intermediate velocity field into the continuity equation (\ref{EQ.1}) to get the pressure Poisson equation, and solving it to obtain the second intermediate velocity field; (4) applying an appropriate mean pressure gradient to the second intermediate velocity field with the pressure term to maintain a constant flow rate. 
	 
	A striking highlight in the numerical algorithm is that, the tensor-based interpolation method we proposed (\cite{Zhang23}) is adopted to address the well-known high Weissenberg number problem (HWNP). This method can guarantee the invariants and symmetric positive definite property of the conformation tensor without adding any artificial diffusion, so as to achieve feasible and accurate numerical simulation of high and ultra-high $Wi$ viscoelastic fluid flow. The key of the tensor-based interpolation method is to interpolate the eigenvalues and orientation of the conformation tensor field {\bf C} rather than its components. The brief application steps are as follows:
			
	(i) Decomposing a known conformation tensor field \textbf{C} by $\textbf{C} = \textbf{R} \rm \boldsymbol{\Lambda} \textbf{R}^T$ to obtain $\textbf{R}$ and $\boldsymbol{\Lambda}$;
	
	(ii) Calculating the Eulerian angles and eigenvalues at the grid node $i$ by (\ref{EQ.5}) and (\ref{EQ.6}),
	\begin{equation}
	\rm \boldsymbol{\Lambda} = \begin{bmatrix}
		\Lambda_1 & 0 & 0 \\
		0 & \Lambda_2 & 0 \\
		0 & 0 & \Lambda_3 \\
	\end{bmatrix}\label{EQ.5},
    \end{equation}
    \begin{equation}
	\textbf{R} = \begin{bmatrix}
		\cos\theta \cos\varphi & \sin\psi \sin\theta \cos\varphi-\cos\psi \sin\varphi & \cos\psi \sin\theta \cos\varphi+\sin\psi \sin\varphi \\
		\cos\theta \cos\varphi & \sin\psi \sin\theta \sin\varphi+\cos\psi \cos\varphi & \cos\psi \sin\theta \sin\varphi-\sin\psi \cos\varphi \\
		-\sin\theta & \sin\psi \cos\theta & \cos\psi \cos\theta \\
	\end{bmatrix} \label{EQ.6},
    \end{equation}
    where \textbf{R} and $\bf{\Lambda}$ are the rotation and the diagonal matrix, respectively; $\psi$, $\theta$ and $\varphi$ the Euler angles relative to the Cartesian coordinate system; $\lambda_{1}$, $\lambda_{2}$ and $\lambda_{3}$ are the eigenvalues;
	
	(iii) Obtaining the Eulerian angles and eigenvalues at the grid interface \textit{i}+1/2 through various interpolation schemes, and calculating $\boldsymbol{\Lambda}_{i+1/2}$ and $\textbf{R}_{i+1/2}$ at the grid interface by (\ref{EQ.5}) and (\ref{EQ.6}) to reconstruct the conformation tensor $\textbf{C}_{i+1/2}$.
	
	In addition, various curve interpolation schemes can be used to obtain the Euler angles and eigenvalues on the grid interface, and then the conformation tensor at the grid interface can be obtained. So the convection term in equation \ref{EQ.5} adopts the Weighted Essentially Non-Oscillatory (WENO) scheme (\cite{Shu98}). The details can be found in \cite{Zhang21a,Zhang23} and \cite{Zhang24}.

	\subsection {Numerical conditions}
	
	Our latest work (\cite{Zhang24}) demonstrated that the size of the computational domain has a significant effect on the achieved flow state of viscoelastic fluid. And even the occurrence of the steady arrowhead regime (SAR) state with low resistance is related with the shorter channel length. Only the simulation with a sufficiently long channel length can guarantee a fully developed EIT at high $Wi$. Moreover, the desired computational domain length of the simulated case has a functional relationship with $Wi$. Therefore, we use the sufficiently long channel ($L_x \times L_y = 20h \times 2h$) to conduct long-time numerical simulations and statistical analysis. The grid is set to be uniform in streamwise direction and non-uniform in wall-normal direction to ensure that there are more dense grids near the wall. The grid distribution along the wall-normal direction is as follows: 
	\begin{equation}
	{y_j = \frac{1}{a}\tanh(\frac{1}{2}\delta_j \ln(\frac{1+a}{1-a}))}\label{EQ.7},
    \end{equation}
    where $y_j$ is the location of the $j$-th grid node, $\delta_j=-1+2j/{N_y}$ with $N_y$  the total number of grids along the wall-normal direction, $a$ is the regulatory parameter for the grid inequality degree and set to be $a = 0.95$. After the detailed grid independence validation in our recent work (\cite{Zhang24}), the grid resolution is set to $N_x \times N_y = 1024 \times 304$ and the time step is $5\times10^{-4}h/u_b$ or the smaller.
	Table \ref{tab:table1} shows the numerical conditions and some results of DNSs in this study. All simulations use the parameters same as the setup in \cite{Zhang24} fixing $Re = 2000$, $\beta = 0.9$ and $L = 100$. A wide range of $Wi$ is covered from 2 to 200. Also, it is observed that the flow states of $Wi \textgreater 8.8$ are EIT. The observation is consistent with the results of \cite{Shekar20} under the same $Re$. Notably, the characteristic velocity scaling differs between studies: while \cite{Shekar20} employs the Newtonian laminar centerline velocity ($U_c$) , our study uses the bulk flow velocity ($U_b$). This scaling difference creates a systematic offset, where our $Wi = 10$ case corresponds to $Wi_c= 15$ in \cite{Shekar20}. Subsequent analysis focuses on the cases of $Wi \geq 10$ which captures the developed EIT dynamics.

	\begin{table}
	\centering
	\begin{tabular}{ccccccccccccccc}
		
		$Wi$ & $Re$ & $L$ & $\beta$ & $C_{f,tur} \times 10^{-3}$ & $C_{f,lam} \times 10^{-3}$& Flow state \\
		\hline
		2 &2000 &100 & 0.9& / & 3.00 & L \\
		8.8 &2000 &100 & 0.9& / & 2.98 & L \\
		9 &2000 &100 & 0.9& 2.99 & / & EIT \\
		10 &2000 &100 & 0.9& 3.27 & / & EIT \\
		20 &2000 &100 & 0.9& 3.97 & / & EIT \\
		40 &2000 &100 & 0.9& 4.45 & / & EIT \\
		60 &2000 &100 & 0.9& 4.75 & / & EIT \\
		80 &2000 &100 & 0.9& 4.96 & / & EIT \\
		100 &2000 &100 & 0.9& 5.10 & / & EIT \\
		150 &2000 &100 & 0.9& 5.36 & / & EIT \\
		200 &2000 &100 & 0.9& 5.58 & / & EIT \\
	\end{tabular}
	\caption{\label{tab:table1} Numerical conditions and some results. ``L" denotes the laminar flow;  $C_{f,lam}$ represents the friction factor of viscoelastic laminar flow.}	
\end{table}

	\section {Results and analysis}\label{3}
	\subsection {Statistical analysis}
		
\begin{figure}
	\centering
	\includegraphics[width=0.7\textwidth]{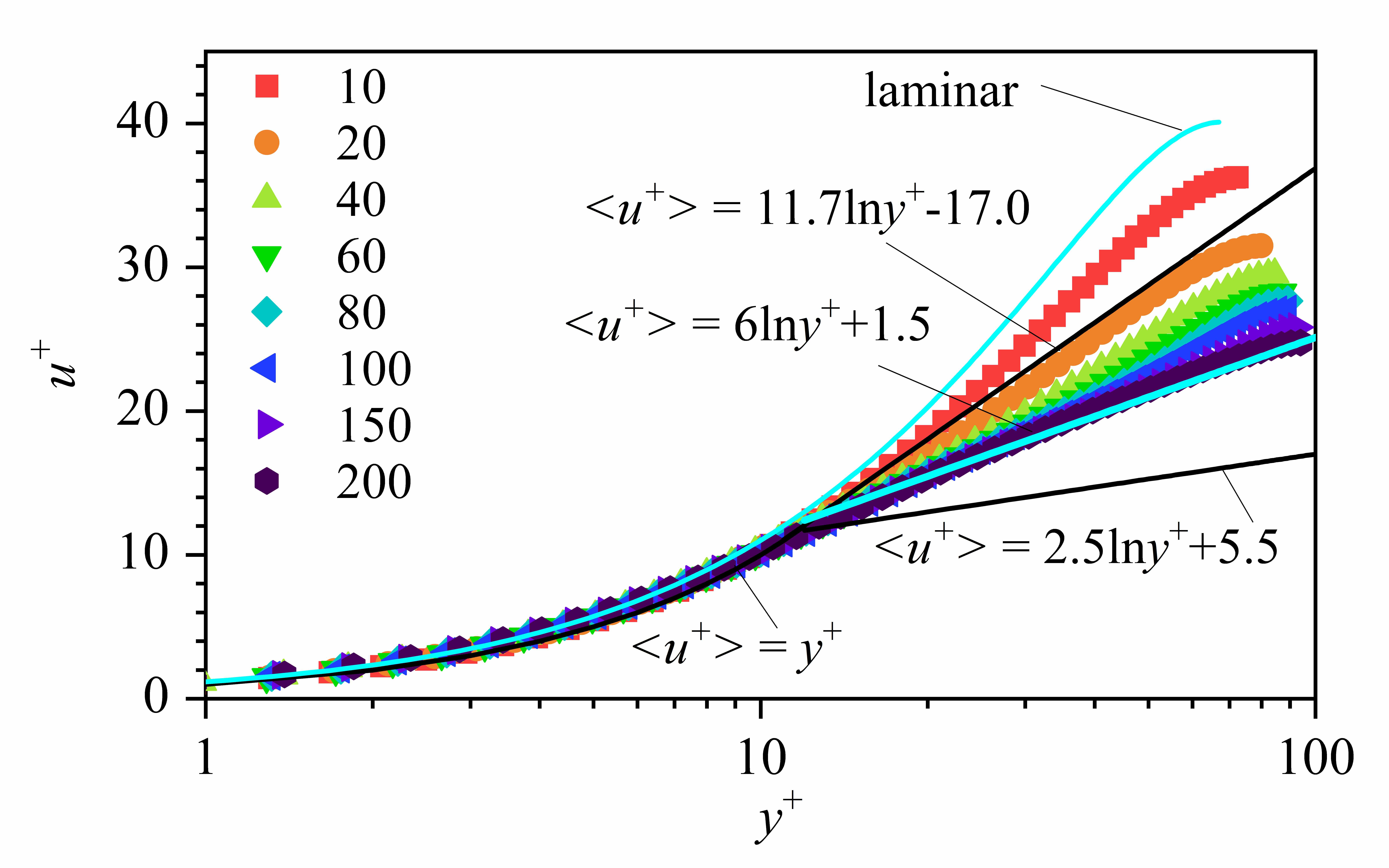}
	\caption{\label{fig1} Mean velocity profiles normalized by the inner scale under different Wi. $u^+$ and $y^+$ are the dimensionless parameters normalized by the skin-friction velocity $u_{\tau} = \sqrt{\tau_w/\rho}$ ($\tau_w$ is wall shear stress) and the zero-shear-rate kinematic viscosity $\nu = \mu/\rho $ of the solution.}
\end{figure}

	We begin by presenting key findings from our flow statistical analysis. Figure \ref{fig1} illustrates  the profiles of mean streamwise velocity normalized by the inner scale across various Wi. Previous investigations (\cite{Dubief13}; \cite{Zhang21a,Zhang21b}; \cite{Wang23}) have established that 3D flows exhibit a systematic upward shift in velocity profiles with enhanced elasticity, ultimately approaching the MDR asymptote (\cite{Virk1970}, $u^+ = 11.7 \ln y^+ - 17.0$)  from the IT baseline ($u^+ = 2.5 \ln y^+ + 5.5$).  In stark contrast, our 2D flow results reveal an inverse behavior: velocity profiles progressively deviate downward from the laminar state with increasing $Wi$, ultimately converging to a distinct asymptotic regime ($u^+ = 6 \ln y^+ + 1.5$) when $Wi$ = 200. This counterintuitive trend aligns with recent friction factor observations (\cite{Zhu21}), collectively suggesting fundamentally different underlying mechanisms between dimensional configurations.  This dimensional dichotomy stems from contrasting dynamical frameworks. In 2D flows, turbulence generation is exclusively governed by elasto-inertial instability (\cite{Sid18}; \cite{Zhu21}), resulting in pure EIT devoid of IT components. Enhanced elasticity amplify EIT intensity, driving the velocity profile's steady transition from laminar to turbulent characteristics. Conversely, 3D systems exhibit dual-regime turbulence (\cite{Wang23}), where EIT and IT dynamics coexist in a competitive interplay. The dominance shift from IT to EIT with increasing Wi suppresses inertial contributions, ultimately steering the system toward MDR asymptotics.  Notably, while 3D flows demonstrate progressive IT suppression and EIT ascendancy, 2D systems manifest unadulterated EIT intensification. This fundamental distinction inevitably produces divergent elastic responses in statistical flow descriptors.
Regarding asymptotic behavior, neither the MDR nor IT asymptote appears to influence velocity convergence in 2D EIT systems. Furthermore, we hypothesize that the emergent asymptotic state may exhibit parametric sensitivity to variations in in $Re$, $\beta$, and $L^2$, suggesting potential avenues for future investigation.
	
\begin{figure}
	\centering
	\includegraphics[width=0.7\textwidth]{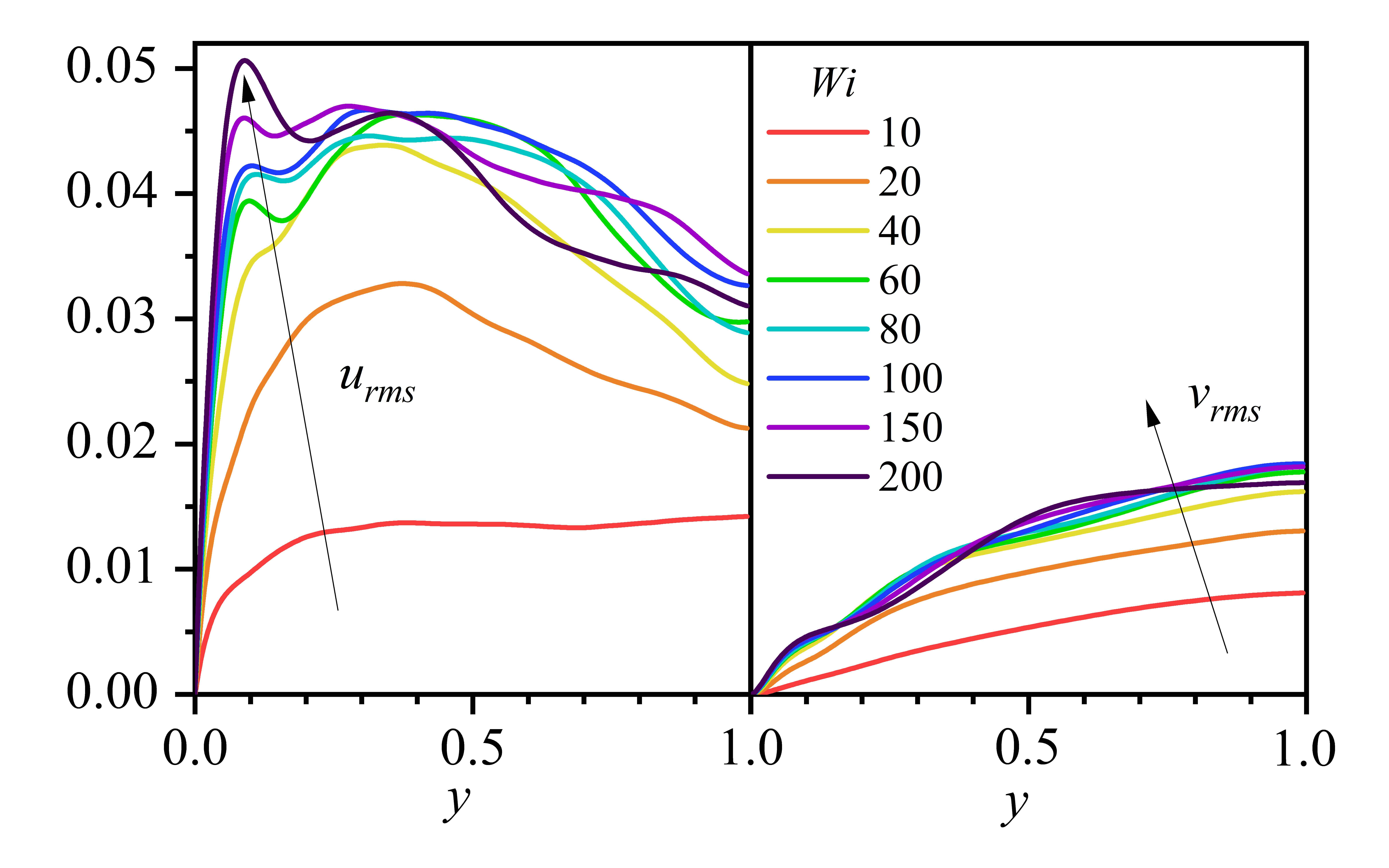}
	\caption{\label{fig2} Distributions of r.m.s. of velocity fluctuations at different Wi.}
\end{figure}

	Figure \ref{fig2} shows the distributions of root mean square (rms) of the  streamwise velocity $u_{rms}$ and wall-normal velocity $v_{rms}$ at different Wi. As is shown, $u_{rms}$ and $v_{rms}$ both tend to increase with Wi and eventually converge, which is similar to the trend observed in Figure \ref{fig1} and opposite to those of the 3D flows (\cite{Wang23}). This still reflects the different dynamical processes in 2D and 3D flows.	 For $u_{rms}$, the peak is increasingly larger, and gradually moves towards the channel wall with the increase of Wi. This indicates increasing velocity fluctuations near the wall and the existence of strong extension structures. For $v_{rms}$, the largest values are at the channel centre, which is consistent with that of 3D EIT with high Wi (\cite{Dubief13}; \cite{Wang23}). In fact, for IT and DRT before the maximum drag reduction, the peak of $v_{rms}$ profile appears far from the centre, which can be considered as a feature that distinguishes EIT from IT and non-MDR states. Moreover, the $v_{rms}$ profile has apparently smaller value than the $u_{rms}$ profile and nearly converges when $Wi \textgreater $ 60, indicating the significant anisotropy of EIT.
		
In channel flow, the total flow drag can be composed of the Reynolds stress $\tau_R$, elastic stress $\tau_e$, and viscous stress $\tau_v$, governed by the stress balance equation:
\begin{equation}
    	{\underbrace{-\overline{u^{\prime}\nu^{\prime}}}_{\tau_{R}}+\underbrace{\overline{\tau_{xy}}}_{\tau_{e}}+\underbrace{\frac{\beta}{\text{Re}}\frac{d\overline{u}}{dy}}_{\tau_{v}}=\tau_{total},}\label{EQ.3-1}
\end{equation}
    where $\overline{(\cdot)}$ represents the ensemblly-averaged variable; $(\cdot)^{\prime}$ is the fluctuating variable and $(\cdot)^{\prime} = (\cdot)-\overline{(\cdot)}$.     

\begin{figure}
    \centering
    \includegraphics[width=0.5\textwidth]{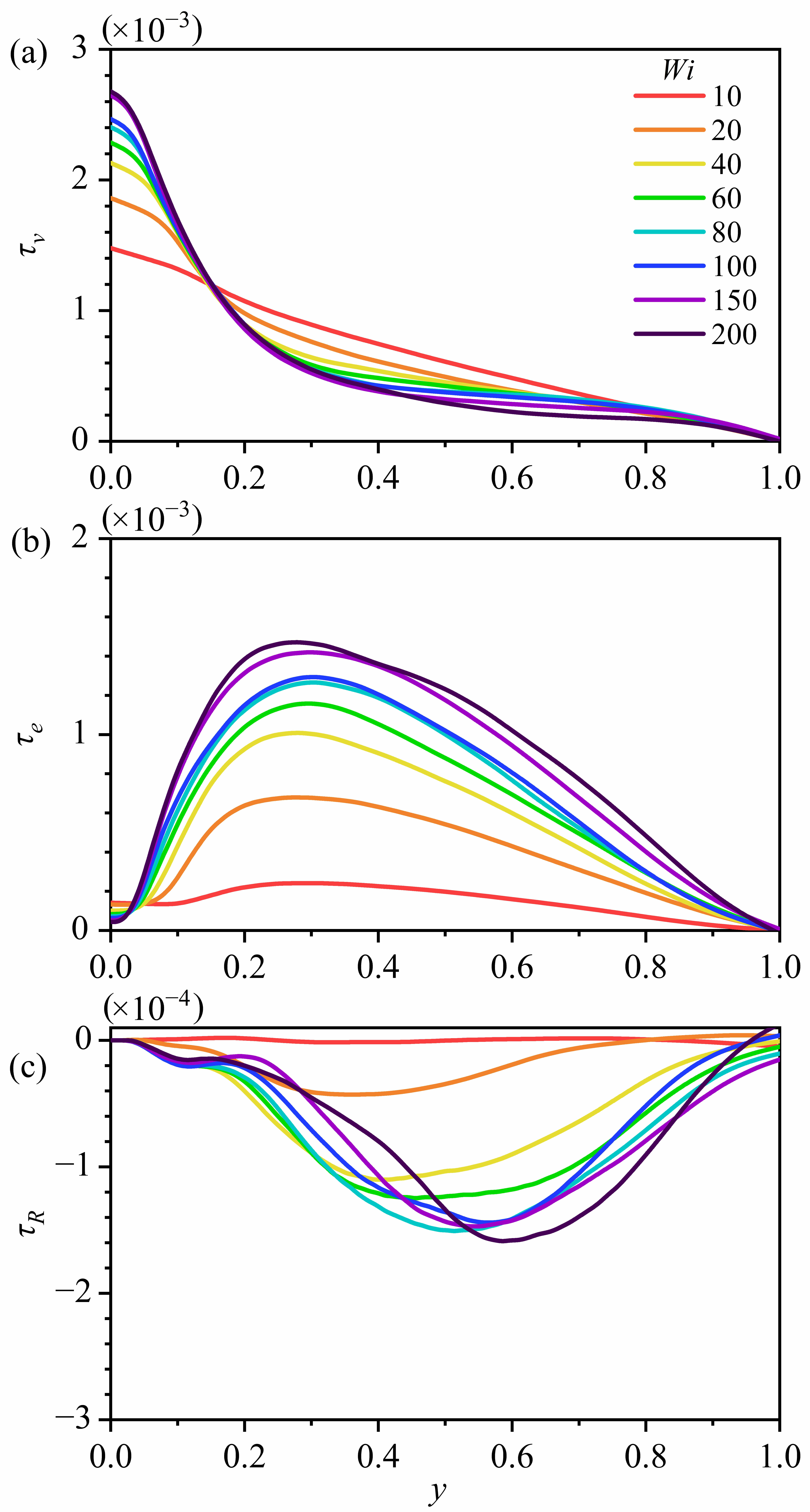}
    \caption{\label{fig3} {Distributions of three kinds of stress at different Wi.}}
\end{figure}

Figure \ref{fig3} illustrates the distributions of mean $\tau_v$, $\tau_e$ and $\tau_R$ at different Wi. Figure \ref{fig3}a reveals a non-monotonic trend in  $\tau_v$ , initially increasing before diminishing beyond a critical y-position, ultimately converging toward the channel center. Near the wall, enhanced elasticity amplifies the viscous stress contribution to flow resistance, though this effect becomes negligible near the channel center due to the rapidly decaying shear rate. Figure \ref{fig3}b demonstrates a monotonic increase in $\tau_e$  with rising Wi, with peak magnitudes consistently localized between  $y = 0.2$ and $y = 0.3$. This behavior aligns with 3D flow observations (\cite{Wang23}), reflecting the role of elasticity in intensifying EIT.  However, contrary to the positive $\tau_v$ and $\tau_e$, the $\tau_R$ in Figure \ref{fig3}c consistently exhibits negative value although significantly lower than the other stresses. Its value grows with Wi, eventually approaching convergence. This finding deviates fundamentally from experimental measurements (\cite{Warholic99}; \cite{Samanta13}) and 3D simulations (\cite{Dubief13}; \cite{Wang23}), where Reynolds stress remains positive -albeit minimal- in the MDR state. The anomalous negative 
$\tau_R$  further distinguishes 2D EIT from IT where Reynolds stress dominantly governs total drag. Detailed mechanistic insights into this phenomenon will be discussed in Section 3.2.

\begin{figure}
\centering
\includegraphics[width=0.9\textwidth]{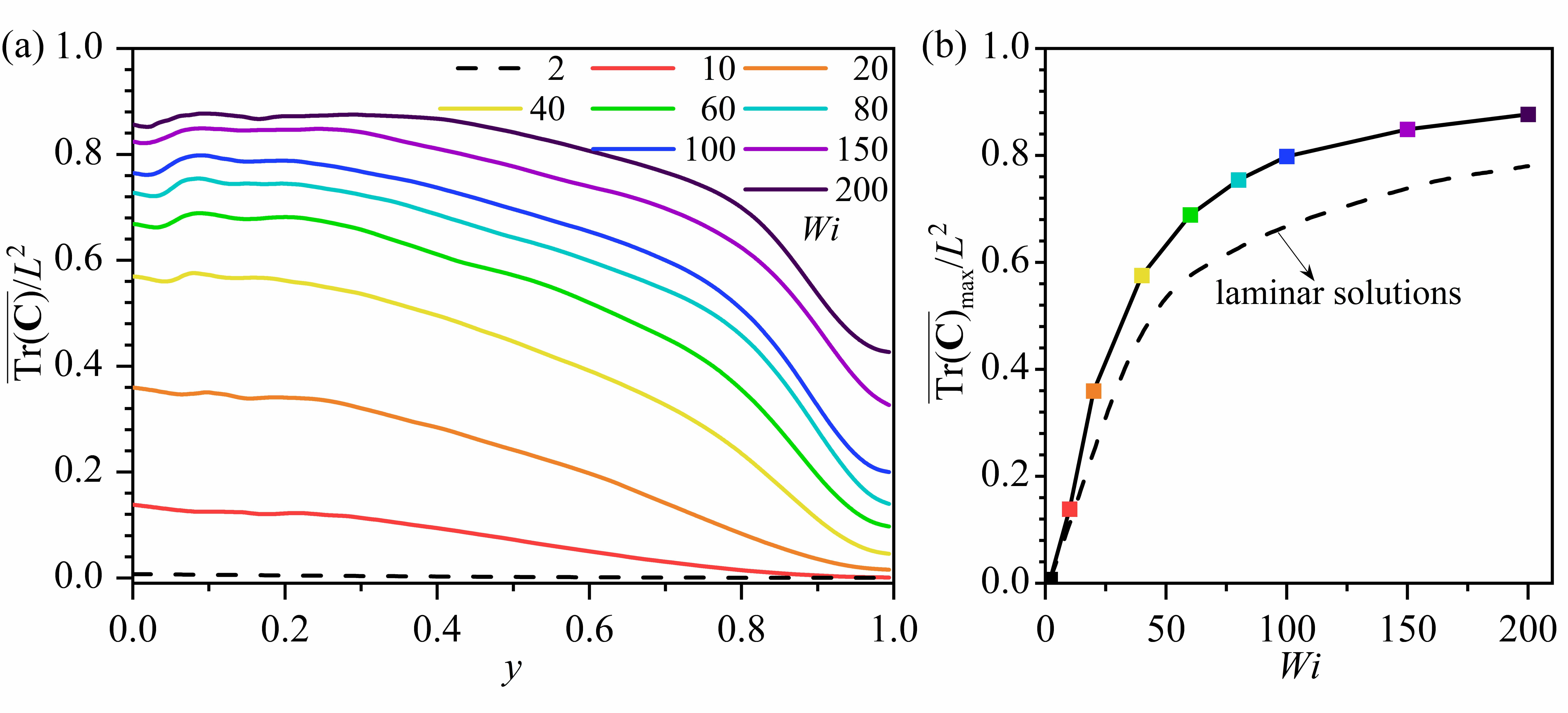}
\caption{\label{fig4} (a) Distributions of mean polymer extension Tr(\textbf{C})/$L^2$ (the $Wi = 2$ case is laminar flow), and (b) the maximum mean polymer extension at different $Wi$ (the dashed line represents the maximum mean polymer extension of corresponding laminar solutions).}
\end{figure}

 Figure \ref{fig4} compares the distributions of mean polymer extension and the maximum relative polymer extension at different Wi. Specifically, as shown in Figure \ref{fig4}a, the extension levels within the central range are consistently lower than those of near-wall range, due to the smaller shear rate away from the wall. However, the extension saturation range has been obviously closer to the centre as Wi increases. In the absence of the suppression by Newtonian turbulence, the polymer extension in 2D EIT can easily reach a high degree as compared to 3D numerical simulations (\cite{Dubief13}). At the same time, compared with the polymer extension in laminar flow ($Wi = 2$), the local peak of turbulent profile away from the wall can be observed, which indicates the existence of the extensional flow topologies described by \cite{Dubief13}. In Figure \ref{fig4}b, by comparing with the maximum mean polymer extension of laminar flow at corresponding Wi, it is observed that EIT can provide an additional amount of extension. The additional extension is randomly distributed in the turbulent extension field (see Figure \ref{fig7}), manifesting as the characteristic polymer extension sheet structures of EIT. Moreover, the promoting effect of increasing elasticity on extension exhibits a saturation trend, which is precisely the embodiment of the effective elasticity (\cite{Zhang24}) we have previously reported. At Wi=200, the maximum mean polymer extension reaches approximately $90 \% $, and will continue to approach slowly $100 \% $ at ultra-high Wi ($e.g., Wi = 1000$, not shown).

	\subsection {Anomalous Reynolds stress}
     
\begin{figure}
	\centering
	\includegraphics[width=1\textwidth]{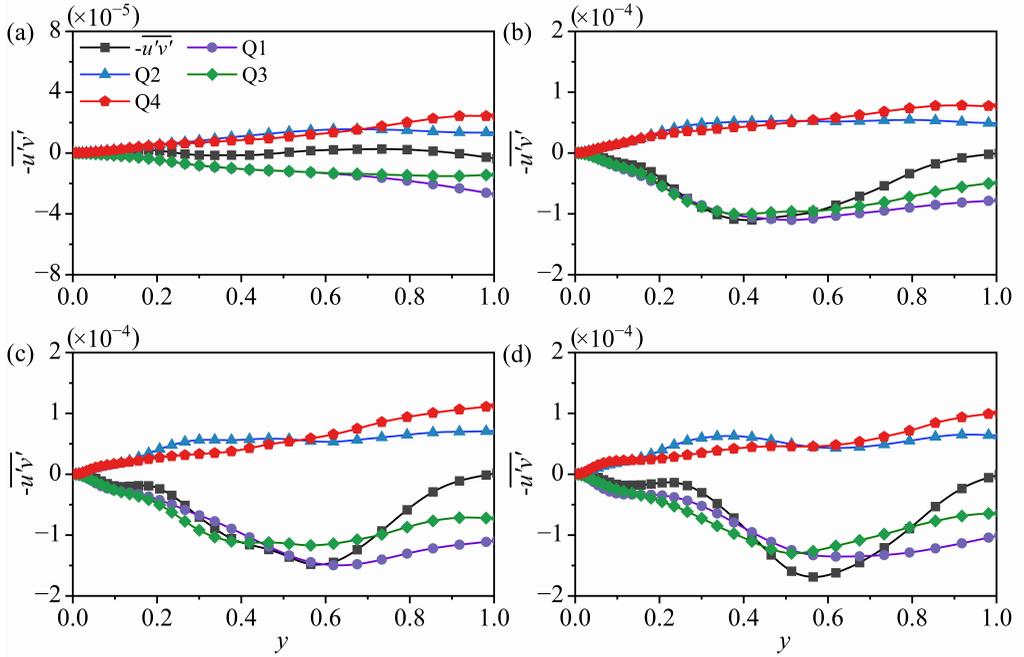}
	\caption{\label{fig5} Quadrant distributions of Reynolds stress at different $Wi$: (a) $Wi$ = 10; (b) $Wi$ = 40; (c) $Wi$ = 100; (d) $Wi$ = 200.}
\end{figure}

     The anomalous Reynolds stress behavior in EIT, as evidenced by Figure~\ref{fig3}c, reveals fundamentally distinct momentum transport mechanisms compared to Newtonian turbulence. The quadrant analysis (\cite{Lu73}) is frequently employed to study the Reynolds stress $\tau_R$. Through quadrant analysis, we systematically deconstruct this phenomenon by resolving the Reynolds stress $\tau_R$ into four components based on the sign of velocity fluctuations $u^{\prime}$ and $v^{\prime}$: Q1 (outward high-speed fluid motion), Q2 (low-speed fluid ejection), Q3 (inward low-speed motion), and Q4 (high-speed sweep motion). In IT, the classical dominance of Q2/Q4 motions drives substantial Reynolds stress production through intense bursting events (\cite{Lu73}). For DRT of viscoelastic fluid, \cite{Cai09} experimentally found that the existence of CTAC additives in viscoelastic fluid can suppress turbulent bursting motions, namely the ejection and sweep motions of Q2 and Q4. This leaded to drag reduction or rather the extremely small Reynolds stress. Similarly, in PAM solution flows, \cite{Guan13} reported that the occurrence of small Reynolds stress is due to the significant decrease of ejections and sweeps of Q2 and Q4 near the wall. Figure~\ref{fig5} reveals two key evolutionary trends with increasing Wi including Q2/Q4 suppression and Q1/Q3 enhancement. The characteristic bursting motions diminish progressively, showing significant reduction in integrated stress contribution. In contrast, symmetric inward/outward motions intensify, generating a mid-channel stress peak ($y \approx 0.4$--0.6) that dominates the anomalous stress profile. The inversion of quadrant dominance at high Wi directly explains the observed negative Reynolds stress. This phenomenon originates from elasticity-driven flow restructuring: polymer stresses amplify symmetric Q1/Q3 motions while suppressing asymmetric bursting events. Crucially, this dual effect persists across both 2D and 3D flow configurations, suggesting a universal EIT mechanism that fundamentally reorganizes turbulent momentum transport pathways. The self-sustaining nature of this regime likely stems from positive feedback between elastic stress accumulation and quadrant motion redistribution \cite{Zhang24}. 

\begin{figure}
	\centering
	\includegraphics[width=1\textwidth]{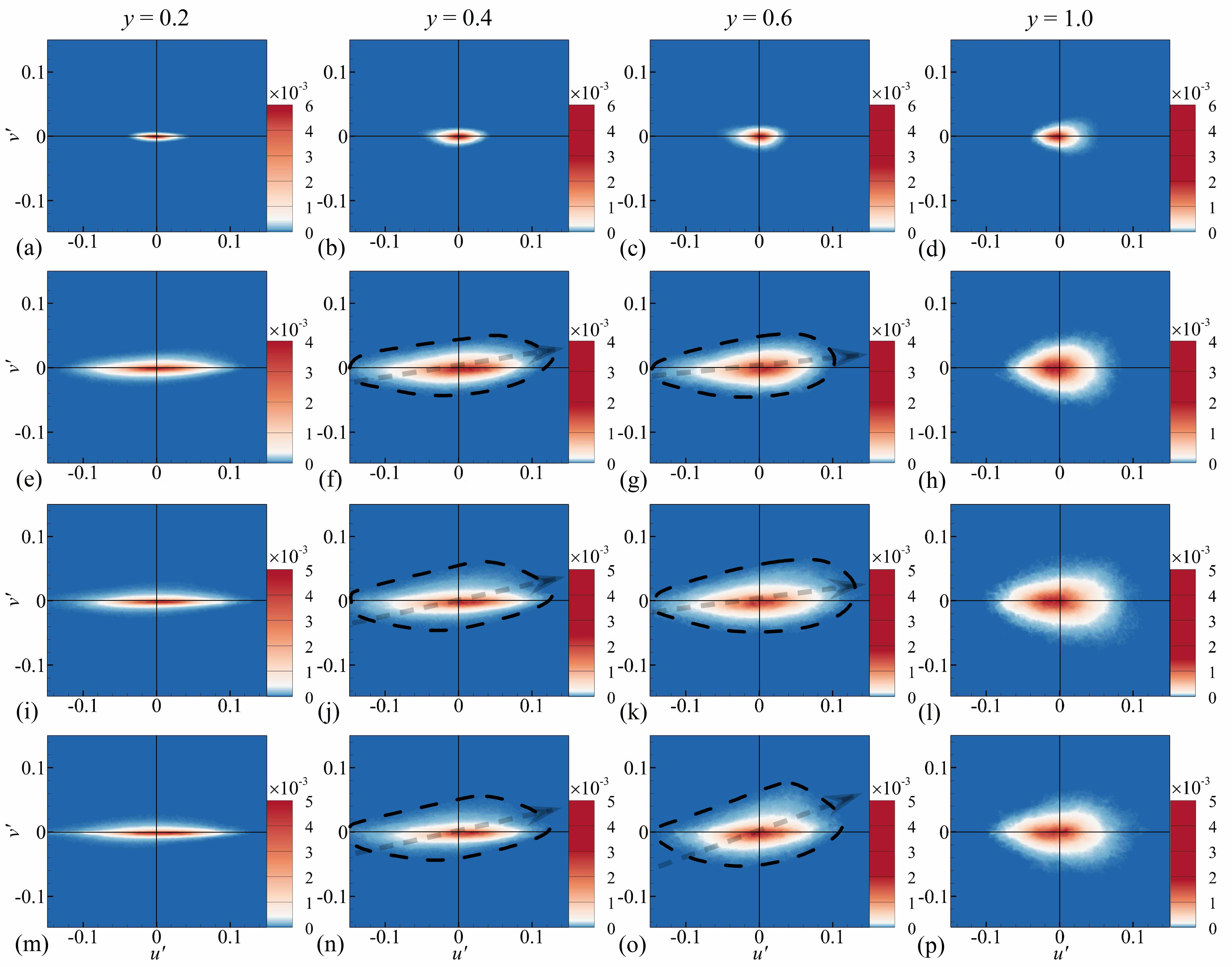}
	\caption{\label{fig6} Probability density distribution contours of four quadrant components at different positions away from the wall (from left to right, $y$ = 0.2, 0.4, 0.6, and 1.0): (a)-(d) $Wi = 10$; (e)-(h) $Wi = 40$; (i)-(l) $Wi = 100$; (m)-(p) $Wi = 200$ (the black dashed frames are used to show the outlines of distribution regions, with the black arrows meaning the oblique distributions).}
\end{figure}

    In order to clearly show the distribution characteristics of four quadrant components, Figure \ref{fig6} presents the relevant probability density distributions in four quadrants at different positions. It is observed that for each Wi case, the distributions of probability density near $(u^{\prime}, v^{\prime}) = (0, 0) $  are the most concentrated, which is the reflection of extremely small Reynolds stress value. Moreover, as the distance from the channel wall increases, the outline of distribution region gradually changes from an elongated strip to an irregular ellipse, indicating that the probability density begins to spread outward (towards the high-value region), especially for $v^{\prime}$. The distribution characteristics are highly similar to the findings in 3D flow experiments (\cite{Guan13}). However, the increasing probability of appearing in high-value region does not necessarily imply a corresponding increase of anomalous Reynolds stress at every point in the flow, due to the mutual restraint and offset among the four quadrants. For example, at $y = 0.2$ , the small Reynolds stress is obviously due to the fact that $v^{\prime}$ has the significantly low-value level. Different from the situation at $y = 0.2$ , at the channel center of $y  = 1.0 $, the probability distributions of high-value region in Q1 and Q4 are quite large. Figure \ref{fig5} shows that at $y = 1.0$ , the values in the upper and lower quadrants almost offset each other, resulting in zero Reynolds stress here. Nonetheless, at $y = 0.4$  and $y = 0.6$, there is a noticeable presence of relatively larger Reynolds stress. As shown by the black dashed frames and arrows, the outlines of distribution regions are obliquely distributed along the first and third quadrants, which is in stark contrast to the diagonal distribution along the second and fourth quadrants in IT (\cite{Guan13}). This indicates that the distributions in Q1 and Q3 are at high probability and high value levels, suppressing the positive contribution of Q2 and Q4 to Reynolds stress, which well explains the location of the peaks of Reynolds stress profiles in Figure \ref{fig5}. Therefore, contrary to the dominance of ejection and sweep motions motions in IT, the above demonstrates the significant dominance of the motions of Q1 and Q3 in 2D EIT.

\begin{figure}
	\centering
	\includegraphics[width=0.9\textwidth]{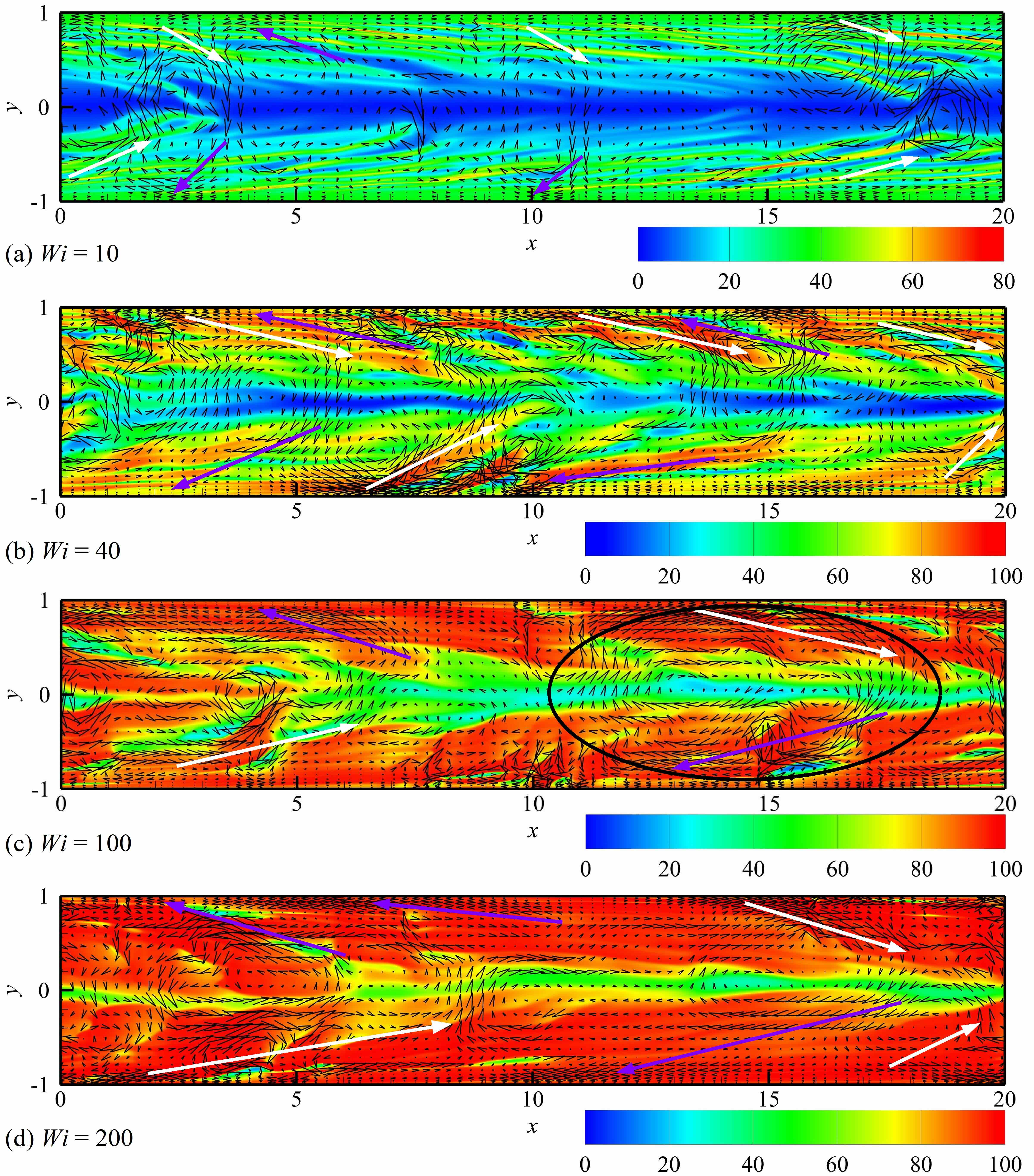}
	\caption{\label{fig7} Superposition of the instantaneous extension and the fluctuating velocity vector field at $Wi$ = 10, 40, 100, and 200 (the $x$ direction is compressed by half for improved display).}
\end{figure}

    Instead of focusing on the frequency and amplitude of the fluctuating velocity at the single position, a more comprehensive understanding can be obtained by analyzing the whole flow field. Figure \ref{fig7} depicts the superposition of instantaneous polymer extension and fluctuating velocity vector field at $Wi$ = 10, 40, 100, and 200. These flow fields are all filled with many vortexes composed of fluctuating velocity vectors, each of which contains the fluctuating velocity vectors of four different quadrant components. The regions with concentrated Q1 and Q3 motions are highlighted by white and purple arrows, respectively. Combined with Figure 6, we can observe that Q1 and Q3 motions are dominant, and this situation becomes more significant as $Wi$ increases. Take the black ellipse region in Figure \ref{fig7}b as an example, which can be approximately regarded as a large vortex. Clearly, the upper and lower parts within this region are undergoing two opposing processes of stretching and contraction of the polymer extension sheet-like structures. In the flow of viscoelastic fluid, the dynamic behavior of polymer chains is intricately linked to the velocity field. At a macro level, the stretching and contraction processes of the polymer extension sheet-like structures must be directly and closely related to the direction of fluctuating velocity vectors. Thus, a clear judgment can be obtained: the vast majority of fluctuating velocity vectors in the first and third quadrants,namely the motions of Q1 and Q3 are arranged along and against the polymer extension sheet-like structures, respectively. For the viscoelastic fluid flow, a typical characteristic is the presence of numerous elongated polymer extension sheet-like structures in the flow field, which exhibit the inclined distribution from the near-wall region towards the channel centre along the streamwise direction. This distribution becomes more pronounced as Wi increases, accompanied by the intensification of the anisotropy of the EIT. Thus, this distribution indicates that the fluctuating velocity vectors accompanying the polymer extension sheet-like structures are preferentially in the first or third quadrant. In 2D EIT, unencumbered by IT, this situation is even more pronounced, leading to anomalous Reynolds stress. Finally, comparing the results of 2D EIT with numerical (\cite{Dubief13}; \cite{Wang23}) and experimental results (\cite{Samanta13}; \cite{Choueiri18}) of 3D turbulent flow, it is evident that the Reynolds stress values in both cases are significantly lower than other stresses. However, it is noteworthy that although they are small enough to be negligible, the Reynolds stress in 3D flow is not negative. We speculate that the presence of residual other dynamics within the MDR state may be a contributing factor, which supports the view that the MDR state is not the pure EIT (\cite{Zhu21}). Nevertheless, the existence of anomalous Reynolds stress indicates that EIT has a strong suppression on IT in the MDR state, and the degree of this suppression can be quantitatively measured by the variation of Reynolds stress. 

	\subsection {Budgets of turbulent kinetic energy and elastic energy}
In the MDR state, the negligible Reynolds stress cannot dominate the production of turbulent kinetic energy (TKE), and polymers provide almost all the energy required for turbulence maintenance (\cite{Min03}; \cite{Dubief13}; \cite{Wang23}). Similarly, in the above section, the discovery of minimal and anomalous Reynolds stress also proves that the main source of TKE in EIT cannot possibly be the production term related to the Reynolds stress. Therefore, the energetic properties in 2D EIT are well worth exploring and compared with those of 3D flow. The transformation between the turbulent elastic energy (TEE) and TKE greatly distinguishes the viscoelastic fluid flow from the Newtonian fluid flow. The integral form of the budget of TKE is as follows:

\begin{equation}
	{\int_0^2{P_k\mathrm{d}y}-\int_0^2{\varepsilon_k\mathrm{d}y}-\int_0^2{G}\mathrm{d}y=0,}\label{EQ.3-2}
\end{equation}
\begin{equation}
	{P_k = -\frac{1}{2} \overline{u^{\prime} v^{\prime}} \frac{\partial \bar{u}}{\partial y}, \quad
		\varepsilon_k = \frac{\beta}{\text{Re}} \overline{\frac{\partial u_i^{\prime}}{\partial x_j} \frac{\partial u_i^{\prime}}{\partial x_j}}, \quad
		G = \overline{\tau_{ij}' \frac{\partial u_i'}{\partial x_j},}}\label{EQ.3-3}
\end{equation}
    where $P_k$ is the production term of TKE, $\varepsilon_k$ is the dissipation term, and $G$ is the energy transfer term between TKE and TEE (the positive value of $G$ indicates the transfer from TKE to TEE, while the negative value signifies the reverse process).

Similarly, the budget of TEE can be obtained by integrating the elastic energy transport equation: 

\begin{equation}
	{\int_0^2P_e\mathrm{d}y-\int_0^2\varepsilon_e\mathrm{d}y+\int_0^2G\mathrm{d}y=0,}\label{EQ.3-4}
\end{equation}

\begin{equation}
	P_e = {\overline{\tau_e}\frac{\partial\overline{u}}{\partial y}}, 
	\varepsilon_e = \frac{\overline{f(\mathrm{tr}(\textbf{C}))\tau_{ii}}}{2Wi}\label{EQ.3-5},
\end{equation}
where $P_e$ is the production term dominated by elastic stress of TEE and $\varepsilon_e$ is the dissipation term.

\begin{figure}
	\centering
	\includegraphics[width=0.7\textwidth]{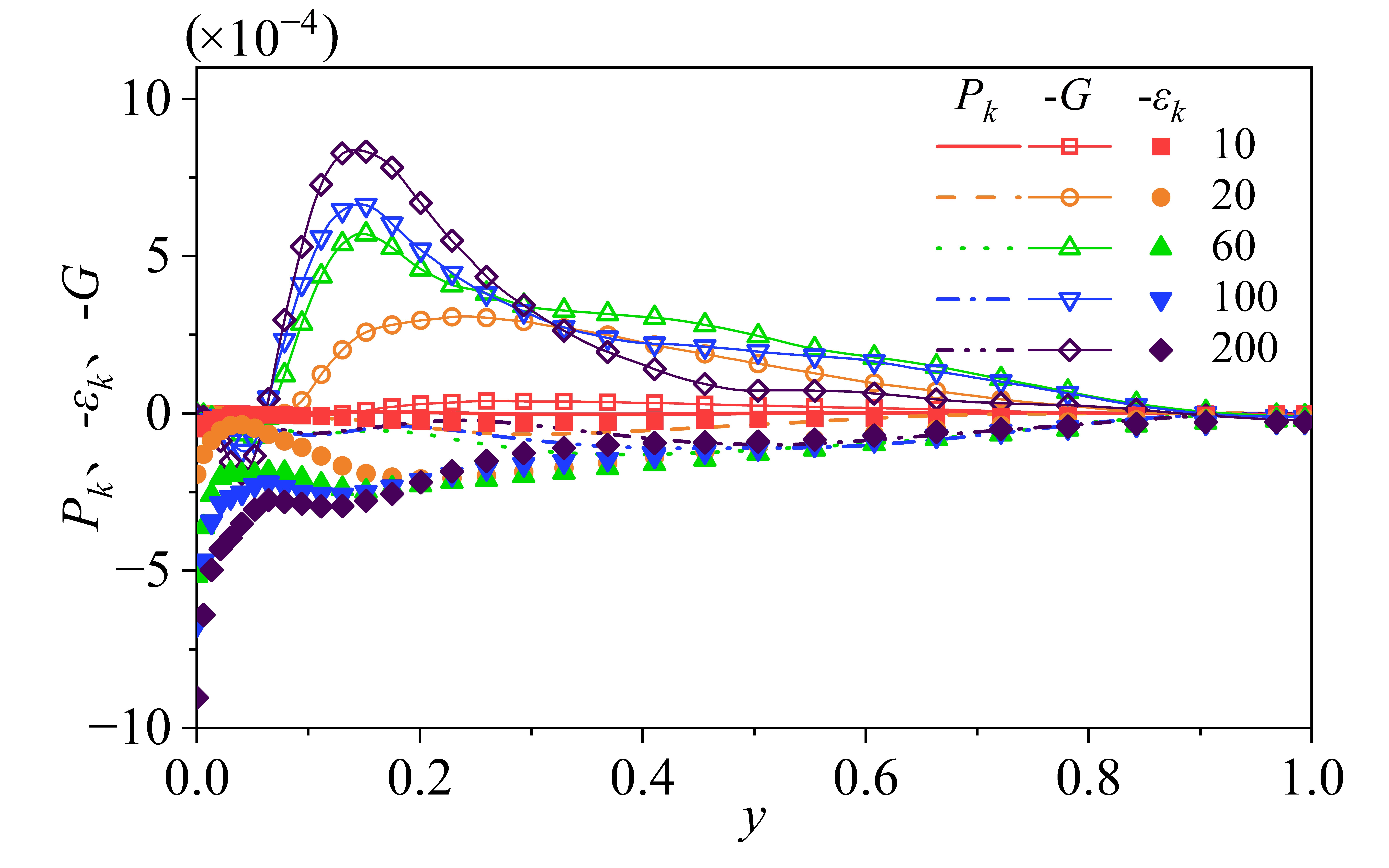}
	\caption{\label{fig8} {Profiles of turbulent kinetic energy budget at different Wi.}}
\end{figure}

    Figure \ref{fig8} illustrates the profiles of the budget of TKE under varying $Wi$. In all cases, the primary source of TKE is consistently the energy transfer term -$G$: TEE is converted into TKE through the interaction between polymers and turbulent flow. Also, this feature is independent of Wi. However, this situation in which the generation of TKE is overwhelmingly dominated by -$G$, only occurs in the asymptotic MDR state or EIT of 3D flow at high Wi (\cite{Dubief13}; \cite{Wang23}). Therefore, we can infer that EIT likely has already formed in actual 3D flow, but its partial features are obscured due to the coexistence with IT. Moreover, -$G$ is reversed within a narrow region near the wall and this phenomenon also exists in 3D flow (\cite{Dubief13}; \cite{Wang23}). The only difference lies in the fact that the production term $P_k$ exhibits a negative value in 2D EIT, which is attributable to the aforementioned abnormal Reynolds stress. Nonetheless, $P_k$ constitutes a small proportion in the budget of TKE, exerting a negligible impact on the self-sustaining mechanism of 2D EIT. 
	
\begin{figure}
	\centering
	\includegraphics[width=0.7\textwidth]{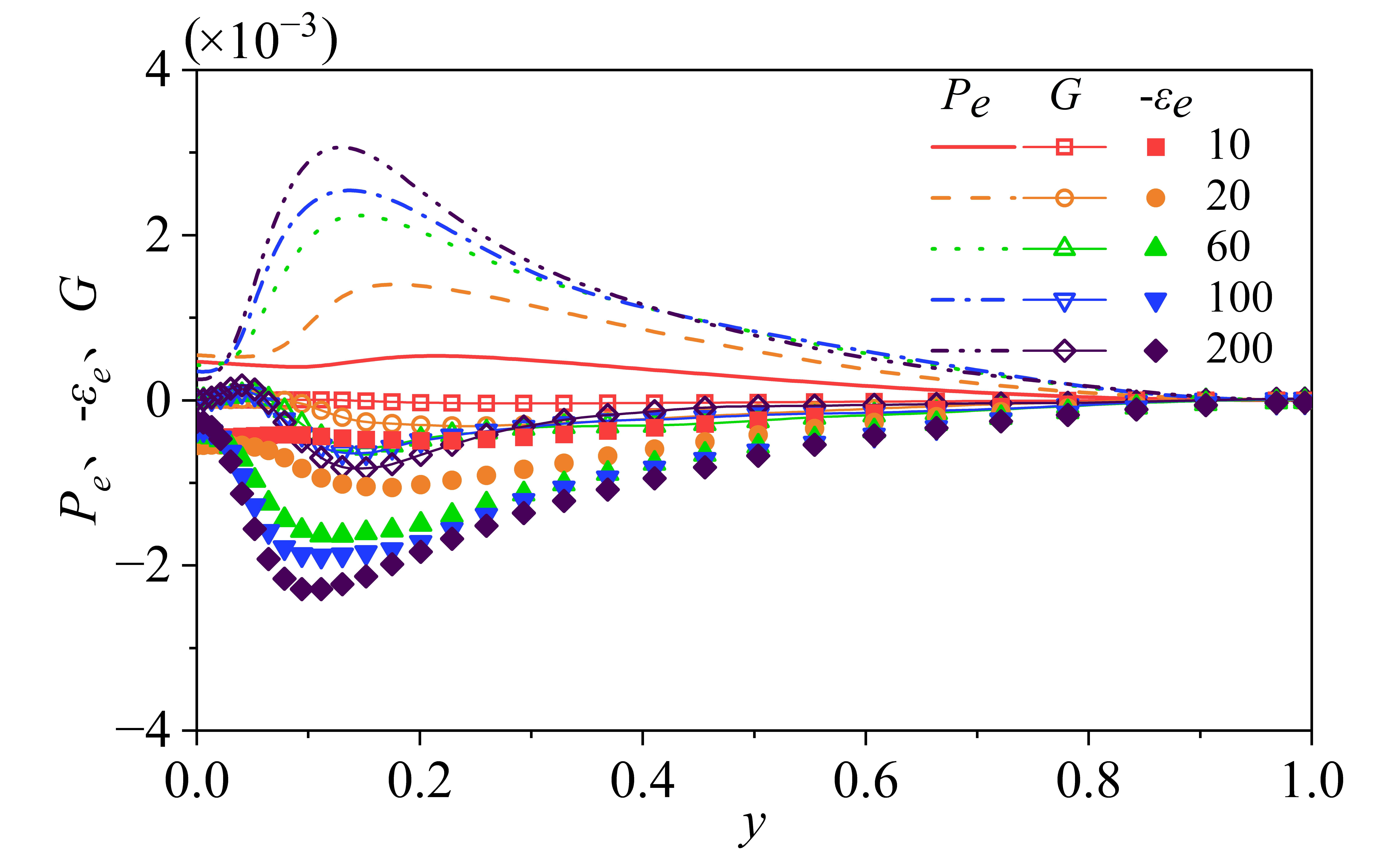}
	\caption{\label{fig9} {Profiles of elastic energy budget at different Wi.}}
\end{figure}

    Similar to Figure \ref{fig8}, Figure \ref{fig9} shows the profiles of the budget of TEE at different $Wi$. The trend of each curve is consistent with the 3D simulation results (\cite{Zhang21b}). The generation and dissipation of TEE are mainly concentrated in the near-wall region, resembling the distribution of the budget of TKE. However, unlike the generation of TKE derived mainly from -$G$, $P_e$ provides the vast majority of the generation of TEE, indicating the dominant role of the work done by the elastic stress. As Wi increases, the enhancement of the turbulent intensity of EIT is primarily driven by the increase of -$G$, yet there is a saturation trend. This is attributed to the concurrent increase of the dominant dissipation term $\varepsilon_e$, resulting in only a smaller fraction of TEE being converted into TKE. The unique TEE characteristics of EIT show remarkable consistency in the budget analysis of between 2D and 3D flow, which is also a strong evidence of the 2D nature of EIT.

	\subsection {Redistributive role of pressure}

    The above content provides a detailed analysis of the TKE budget of EIT. In fact, the majority of studies including this paper choose to use the pressure-driven flow, where the pressure play a critical role in the redistribution of energy among the components of momentum (\cite{Terrapon15}). Moving forward, this paper will decompose the pressure into several parts and assess their roles in redistributing TKE. 
    
    Many scholars have analyzed the redistributive role of pressure in 3D turbulent flow of viscoelastic fluid (\cite{Ptasinski03}; \cite{Terrapon15}; \cite{Zhang22}). As shown in equations (\ref{EQ.3-6}-\ref{EQ.3-12}), the pressure fluctuations can be decomposed into four contributing terms. As reported by Kim et al. (1989) and \cite{Terrapon15}, the rapid part $p^{\prime}_r$ is a direct and linear response to the change and deformation in the mean flow field, while the slow part $p^{\prime}_s$ is the response through nonlinear turbulent interactions. The distinctive polymer contribution $p^{\prime}_p$ exists in the viscoelastic flow due to the polymer stress. Notably, the first two terms ($p^{\prime}_r$ and $p^{\prime}_s$) and $p^{\prime}_p$ are identified as the inertial and elastic contributions, respectively (\cite{Terrapon15}; \cite{Zhang22}). The inertia and nonlinear elasticity of the fluid play varying roles in the pressure redistribution effect and the self-sustaining process of EIT.

	\begin{equation}
	{p^{\prime}(\textbf{x})=p(\textbf{x})-\overline{p}(\textbf{x}),} \label{EQ.3-6}
    \end{equation}

    \begin{equation}
	{p^{\prime}(\textbf{x})=p^{\prime}_{r}(\textbf{x})+p^{\prime}_{s}(\textbf{x})+p^{\prime}_{p}(\textbf{x})+p^{\prime}_{St}(\textbf{x}),} \label{EQ.3-7}
    \end{equation}

    \begin{equation}
	{\nabla^{2}p_{r}^{\prime}=-2\frac{\mathrm{d}\overline{u}}{\mathrm{d}y}\frac{\partial v^{\prime}}{\partial x},} \label{EQ.3-8}
    \end{equation}

    \begin{equation}
	{\nabla^{2}p_{s}^{\prime}=-\frac{\partial u_{i}^{\prime}}{\partial x_{j}}\frac{\partial u_{j}^{\prime}}{\partial x_{i}}+\frac{\mathrm{d}^{2}\overline{{{v^{\prime2}}}}}{\mathrm{d}y^{2}},} \label{EQ.3-9}
    \end{equation}

    \begin{equation}
	{\nabla^{2}p_{p}^{\prime}=\frac{\partial^{2}\tau^{\prime}}{\partial x_{i}\partial x_{j}},} \label{EQ.3-10}
    \end{equation}

    \begin{equation}
	{\nabla^{2}p_{St}^{\prime}=0,} \label{EQ.3-11}
    \end{equation}

    \begin{equation}
	{\Phi_{ii}=2\overline{p^{\prime}\frac{\partial u_i^{\prime}}{\partial x_i}}.}\label{EQ.3-12}
    \end{equation}
    where $p^{\prime}_r, p^{\prime}_s, p^{\prime}_p$, and $p^{\prime}_{St}$ represent rapid, slow, polymer, and Stokes pressures, respectively; and the Stokes pressure $p^{\prime}_{St}$ reflects the role of the wall boundary condition, which is ignored due to its apparently small proportion; $\textbf{x} = (x,y)$, $\otimes^{\prime}$ represents fluctuations and $\overline{\otimes}$ represents average quantities.
    
\begin{figure}
	\centering
	\includegraphics[width=1\textwidth]{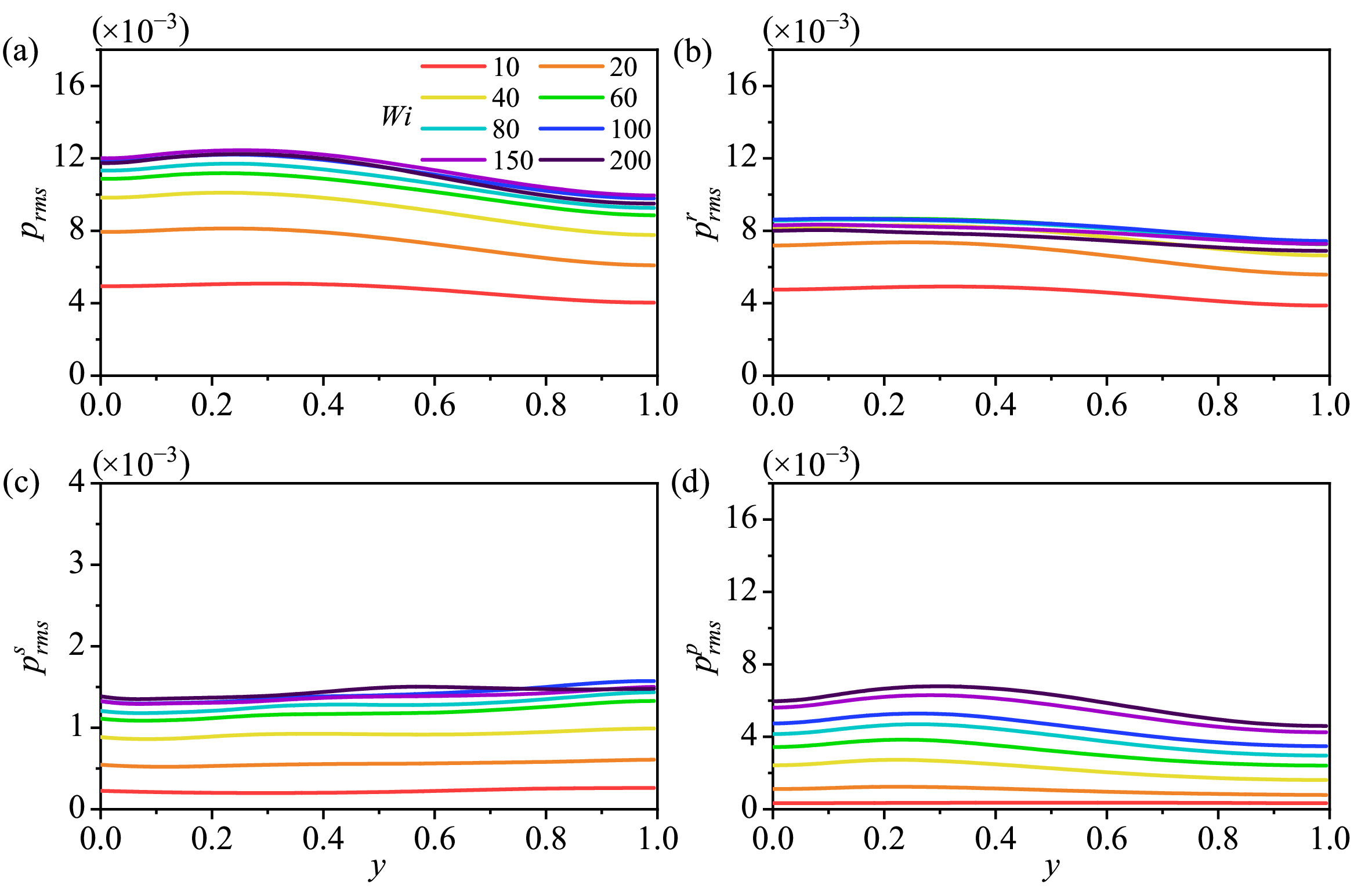}
	\caption{\label{fig10} Pressure r.m.s. distribution along the wall-normal direction with different Wi: total (a), rapid (b), slow (c) and polymeric (d) contribution.}
\end{figure}

\begin{figure}
	\centering
	\includegraphics[width=0.6\textwidth]{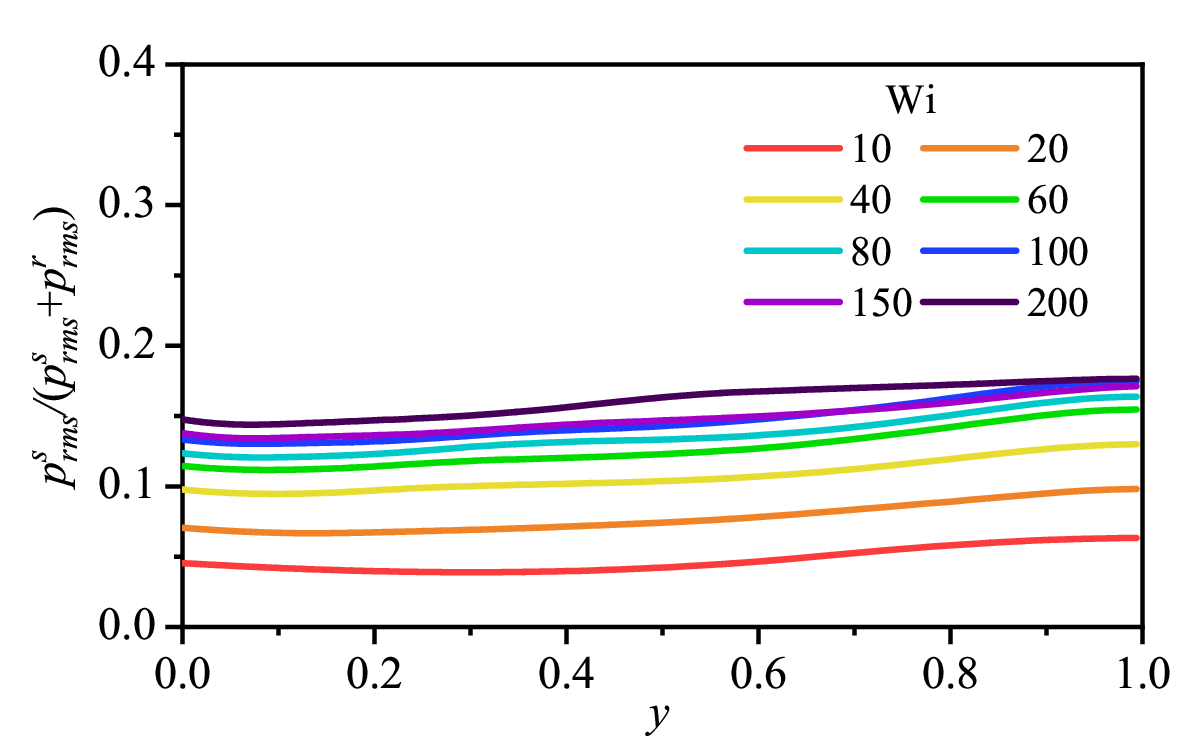}
	\caption{\label{fig11} Distribution of the ratio of $p_{rms}^{S}$ to ($p_{rms}^{R}+p_{rms}^{S}$) along the wall-normal direction.}
\end{figure}

    Figure \ref{fig10} shows the r.m.s. distribution of various pressure contributions at different Wi. As is shown, these different kinds of profiles in 2D EIT are generally similar to the results of the MDR state (\cite{Terrapon15}), characterized by smoothness but not complete flatness. However, in another research with the Oldroyd-B model (\cite{Zhang22}), the r.m.s. profiles of 3D EIT exhibits a quasi-flat distribution. This difference likely arises from the effect of the maximum extension length $L$ ($L$ = $\infty$ for the Oldroyd-B model, $L$ = 100 in this paper and $L$ = 200 in \cite{Terrapon15}).    
    Detailedly, with the increase of $Wi$, $p_{rms}$ gradually increases and approaches convergence, which fits with the effective elasticity effect we proposed (\cite{Zhang24}). Moreover, the $p^{\prime}_s$ r.m.s. values reflect that the contribution of slow pressure is distinctly smaller than the other contributions, in line with the findings by \cite{Terrapon15} in 3D simulations: polymers have a strong suppression on the slow pressure contribution $p^{\prime}_s$. However, for the contribution of rapid pressure $p^r_{rms}$, there exists a decrease trend at high Wi ($Wi \textgreater 80$), which is actually also slightly reflected in the slow pressure. Unlike the effect of relaminarization in the 3D flows with low $Re$ (\cite{Zhang22}), the reason here is the suppression effect of strong elasticity. Furthermore, Figure \ref{fig11} illustrates the distribution of $p^s_{rms}/(p^r_{rms} + p^s_{rms})$ along the wall-normal direction. The increasing trend of this ratio indicates that the contribution of $p^{\prime}_s$ in the inertial pressure gradually approaches saturation. Consequently, the rapid pressure contribution $p^{\prime}_r$ is more significantly suppressed by polymers in 2D EIT. Notably, the visible bump or peak of $p_{rms}$ within $y$ = 0.2-0.4 in Figure \ref{fig10}a, unequivocally originates from the polymer pressure contribution $p^{\prime}_p$, which is in stark contrast to the buffer layer of IT where the obvious peak in the pressure fluctuations stems from the slow pressure (\cite{Terrapon15}). Therefore, especially at high Wi, the increase in total pressure is primarily attributed to the dominant modulating role of polymer pressure $p^{\prime}_p$ in the pressure redistribution effect.

\begin{figure}
	\centering
	\includegraphics[width=1\textwidth]{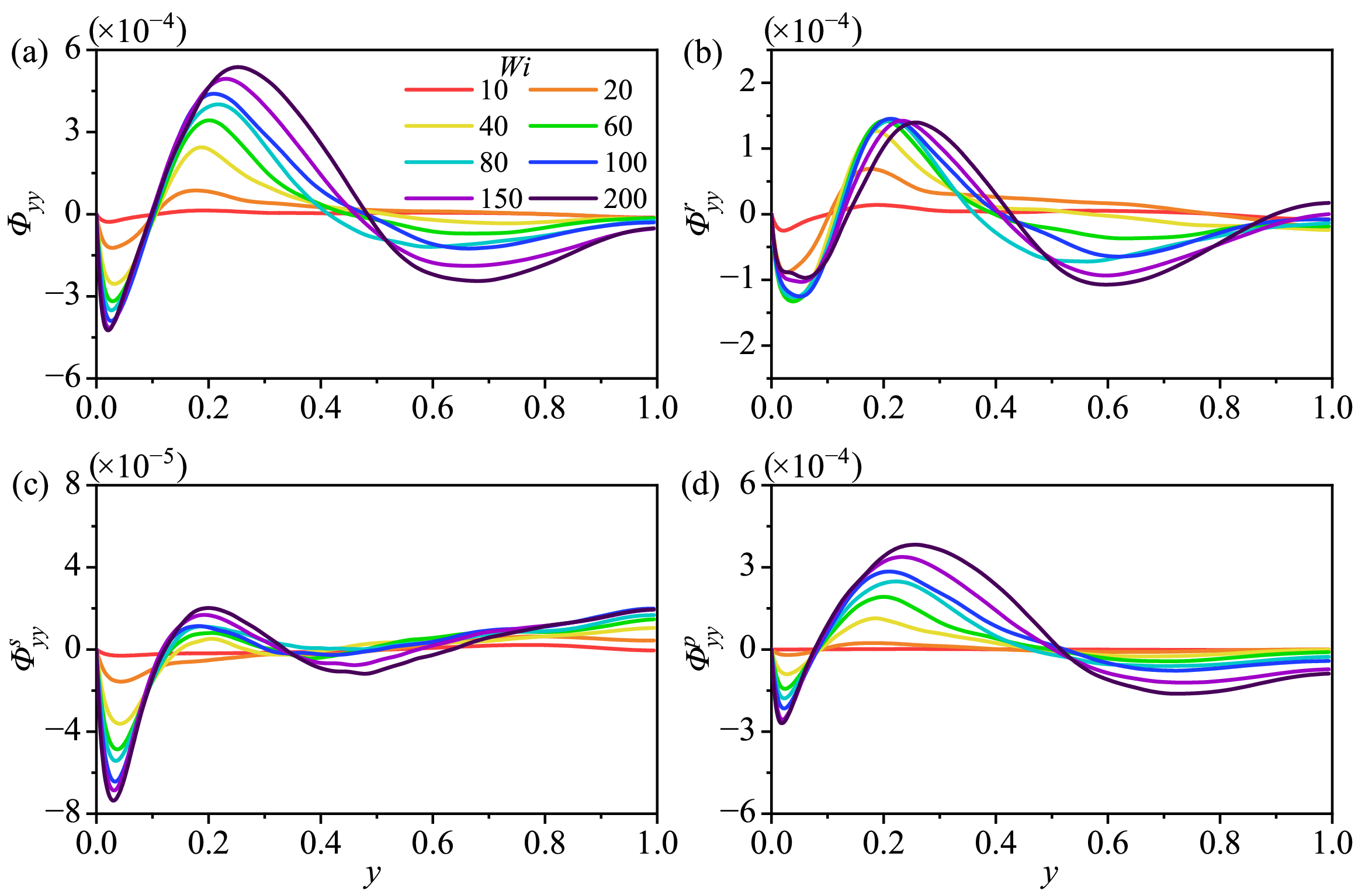}
	\caption{\label{fig12} Distribution of wall-normal pressure-strain component $\phi_{yy}$ from different pressure contributions under different Wi.}
\end{figure}

    According to \cite{Dubief13}, pressure plays a crucial role in redistributing energy, especially TKE, across the momentum components. Through a comprehensive analysis of Reynolds stress transport, \cite{Terrapon15} decomposed the velocity-pressure gradient correlation term into pressure-strain and pressure-diffusion terms, and pointed out that the pressure-strain term $\Phi_{ij}$ transports TKE from the streamwise direction to other directions. The trace of the pressure-strain term has an equality relation tr($\Phi_{ij}$) = $\Phi_{xx} + \Phi_{yy} + \Phi_{zz} = 0$ ($\Phi_{zz}$ is not present in 2D simulations of this paper, and $\Phi_{xx} = - \Phi_{yy}$). Figure \ref{fig12} shows the distribution of the wall-normal pressure-strain component of total, rapid, slow and polymer pressure contributions at different Wi (streamwise pressure-strain component $\Phi_{xx}$ is just the opposite, not shown). Overall, the distribution profiles of different pressure contributions show great consistency with the results of the MDR state as well as 3D EIT (\cite{Terrapon15}; \cite{Zhang22}). Upon comparing the order of magnitude, we can intuitively discern the weak contribution of slow pressure in the energy transport of EIT in Figure \ref{fig12}c. However, unlike the inertial pressure, the polymer pressure contribution does not show an obvious convergence trend due to its elasticity nature, which is also the main source of the continuous enhanced TKE transport effect of pressure. And the $\Phi^p_{yy}$ peak starts to be larger than the $\Phi^r_{yy}$ peak at $Wi \textgreater 40$, which means that the polymer pressure dominates the TKE transport in 2D EIT.
    
    As is shown in Figure \ref{fig12}, at approximately $y \textless 0.1$, all pressure contributions have negative wall-normal pressure-strain component, meaning that the TKE is redistributed from the $y$ direction to $x$ direction. Concurrently, in this range, the negative $-G$ shown in Figure \ref{fig8} indicates the conversion from TKE to TEE, thereby contributing to the characteristic turbulence streaks of viscoelastic flow (\cite{Zhang22}). And the positive $\Phi_{yy}$ at about $0.1 \textless y \textless 0.4$ indicates the formation of wall-normal TKE within this range. Far from the wall (about $0.5 \textless y \textless 1.0$), the formation of streamwise TKE occurs again, with the rapid pressure and polymer pressure dominating this process. In addition, near the channel center, the positive $\Phi^r_{yy}$ and $\Phi^s_{yy}$ indicates that the inertial pressure seems to have a promotion effect on the formation of wall-normal TKE, although offset by the negative effect of the polymer pressure. Notably, in fact, \cite{Terrapon15} found that the polymer pressure contributed more to $\Phi_{yy}$ than $\Phi_{zz}$, in the analysis of 3D flow. They have argued that the elastic contribution promotes the 2D flow to a certain extent, while the inertial part mainly drives the 3D flow. Combined with the dominant role of polymer or elastic pressure on TKE transport of 2D EIT in Figure \ref{fig12}, it is again demonstrated that the 2D nature of elasto-inertial instability dominated by the elasticity in EIT and the objective existence of 2D EIT (\cite{Sid18}).
    
\subsection {Energy spectrum analysis}

\begin{figure}
	\centering
	\includegraphics[width=1\textwidth]{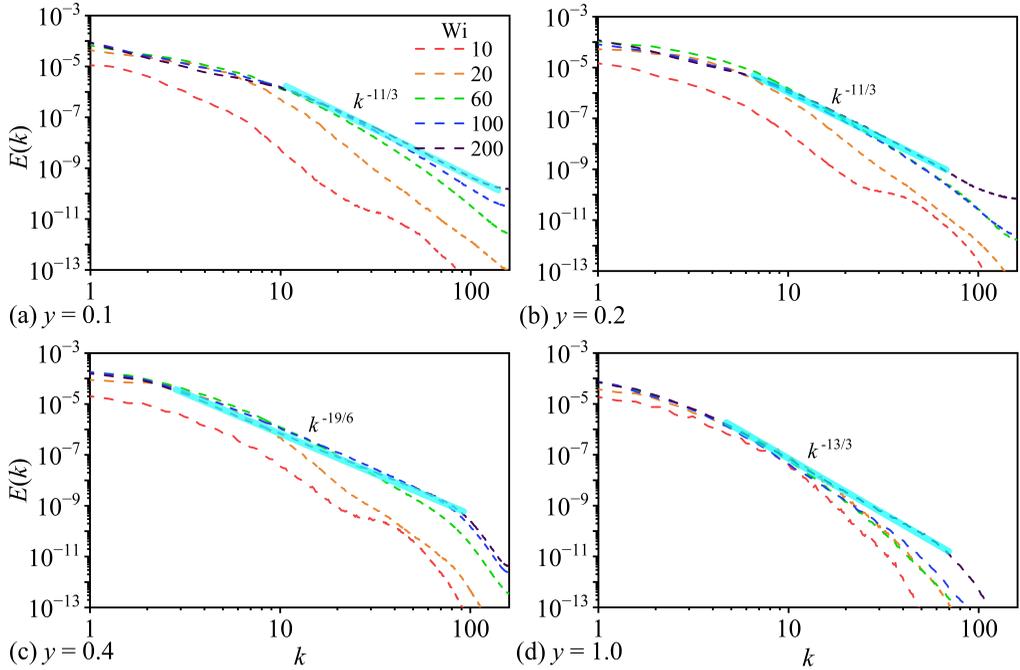}
	\caption{\label{fig13} Energy spectrum at different Wi and wall-normal positions.}
\end{figure}

         From the transformation between TEE and TKE in Figure \ref{fig8} and Figure \ref{fig9}, it can be observed that the energy transformation occurs most intensely within the range of $y$ = 0.1 $\sim$ 0.2, which fully aligns with the characteristic position of energy transformation in 3D numerical simulations (\cite{Wang23}). In order to delve deeper into the dynamics of 2D EIT, an energy spectrum analysis is conducted at this specific range, as well as at $y$ = 0.4 and the centreline of channel. Figure \ref{fig13} shows the energy spectrum of TKE along the streamwise direction under different Wi at different wall-normal positions, with the corresponding power-law spectral decay $k^{-\alpha}$ of interest. For these four different positions, $\alpha$ shows a decreasing trend with Wi, and converges to a specific slope of the line at very high Wi ($e.g.$, $Wi \textgreater 100$). In the near-wall energy spectrum of TKE in Figure \ref{fig13}a and b, the converging power-law decay of 2D EIT is $k^{-11/3}$ within $k \textgreater 10$. Similarly, in 3D EIT, \cite{Dubief13} found the power-law decay at high wavenumber seems to be closer to $k^{-11/3}$ and deviate from $k^{-14/3}$, which is a support point for the 2D EIT. In the energy spectrum of $y$ = 0.4 in Figure \ref{fig13}c, it is observed that the $k^{-19/6}$ power-law decay is present within a large range of $k \approx 4 - 80$, demonstrating the mighty presence of EIT dynamics in this region. However, for the 3D EIT, the $k^{-19/6}$ power-law decay occurs due to the gradual conversion from -5/3 power-law decay with the enhanced elasticity in the low wavenumber region (\cite{Zhang21a}). Its appearence represents the prominent presence of EIT dynamics that gradually replaces IT dynamics. In the energy spectrum of the channel center in Figure \ref{fig13}d, there is also a special power law with -13/3 at $k \approx 5 - 60$, which is between -11/3 and -14/3 and fits the power-law decay of 3D EIT (\cite{Dubief13}; \cite{Steinberg21}). 
         
         \cite{Fouxon03} concluded that $\alpha \textgreater 3$ through theoretical derivation, which is recognized as the theoretical power-law decay of ET (\cite{Steinberg21}). In contrast, the spectral properties of EIT have been poorly studied, but the $k^{-\alpha}$ power-law decay with $\alpha \textgreater 3$ for ET is also proved to be an available feature of EIT in experiments (\cite{Yamani21}). Anyhow, all the above power-law decays for 2D as well as 3D EIT satisfy $\alpha \textgreater 3$. More strikingly, the trend of $\alpha$ decreasing with Wi means that EIT is likely to have a close link to ET. In fact, under the same parameters, the $\alpha$ for 3D flow exhibits a significant difference compared to that of 2D EIT (\cite{Sid18}), which means that the convergence of power-law decay for 2D EIT requires higher Wi. It is more noteworthy that this also shows that the quantitative analysis of both 2D and 3D EIT is likely to encounter unsatisfactory situations, so the qualitative analysis is necessary. Therefore, to some extent, we can still find qualitatively the similarities between 2D and 3D EIT with high Wi in terms of the power-law decay, consistent with the viewpoint of \cite{Sid18}.

	\section {Concluding remarks}\label{4}

    Based on the DNSs of 2D sufficiently long channel flow of viscoelastic fluid, this paper qualitatively analyzes the statistical characteristics and underlying dynamic mechanisms of 2D EIT by compared with 3D flow.
    
    (1) Firstly, the effects of elasticity on the statistical characteristics in 2D flow exhibit opposite trends compared with those of 3D flow. This is due to that, as Wi increases, IT is gradually suppressed in 3D flow, while EIT is gradually promoted in 2D flow. These two diverse processes of dynamic evolution inevitably mean opposite responses to elasticity variation.
    
    (2) We have identified the anomalous Reynolds stress that contributes negatively to the flow resistance in 2D EIT. By quadrant analysis, it demonstrates that the motions in the first and third quadrants are dominant. And these motions are closely associated with the polymer sheet-like extension structures. Strikingly, they are inclined from the near-wall region towards the channel centre along the streamwise direction, which coincides with the distribution characteristics of the motions in the first and third quadrants. We believe that the existence of anomalous Reynolds stress is inextricably linked to the formation of the typical structure of EIT.
    
    (3) Finally, through the above analyses, such as the energy budget, redistributive role of pressure, energy spectrum characteristics and so on, we obtain the dynamic mechanisms of 2D EIT. In detail, TEE is almost entirely generated by elastic stress, which is then partly converted into TKE through the energy transfer term; subsequently, TKE is redistributed to different directions by the pressure, especially the critical and even dominant polymer pressure contribution, and The direction of redistribution differs in spatial location; lastly, over a wide high-wavenumber and small-scale range, the decay of TKE follows the unique power-law spectral decay $ k^{-\alpha} (\alpha > 3)$ of viscoelastic turbulent flows. The sufficient similarities to those of 3D EIT convincingly confirms the objective existence of the 2D nature of EIT.
    
    The exploration of 2D EIT still needs to be expanded and deepened. Similar to the study on altering Wi herein, the effects of other parameters such as Re and concentration also remain to be explored. Particularly, the link between EIT and ET under small Re will be highly anticipated. Looking further ahead, by conducting synchronized studies that encompass experiments as well as direct numerical simulations in both two and three dimensions, the future focus will be on unraveling the transition, in-depth mechanism and control of EIT.

	\section*{Acknowledgements}
	This research was funded by the National Natural Science Foundation of China (NSFC 52006249, 12202308, 12472255).
	
	\section*{Declaration of interests}
	The authors report no conflict of interest.
	
	
	\bibliographystyle{jfm}

	
\end{document}